\documentclass[aps,twocolumn,groupedaddress,floatfix,showpacs]{revtex4}
\usepackage{graphics}%

\begin{document}

\bibliographystyle{apsrev}

\title{Transit times of radiation through photonic bandgap materials:
Pulse reshaping of classical and quantum fields in the time domain}

\author{Martin Ligare}
\email[]{mligare@bucknell.edu}
\affiliation{Department of Physics, Bucknell University, Lewisburg, PA  17837}
\author{Aaron Gee-Clough}
\affiliation{Department of Physics and Astronomy, University College London, 
London, England WC1E 6BT}
\altaffiliation[]{Current Address: 102 Parkside Rd,
Silver Spring, MD, 20910}
\author{Charles Doersam}
\affiliation{Department of Physics, Lycoming College, Williamsport, PA 17701}

%\date{\today}

\begin{abstract}
We study the propagation of model electromagnetic pulses through
photonic bandgap materials and relate time-domain pulse reshaping to
observable transit times. For layered dielectric mirrors we
demonstrate how pulse reshaping of slowly varying classical fields
results in transit-time delays that are equivalent to the group delay.
The time-domain analysis emphasizes the causal origin of negative
group delays.  We also study an analogous fully-quantized model and
show how the same time-domain analysis may be used to interpret
observed delays of single-photon fields.
\end{abstract}

\maketitle 

\section{Introduction}

The propagation speed of particles through tunneling regions has been
studied theoretically since the early days of quantum mechanics, and
optical analogs of tunneling have been explored in recent years
because of their experimental accessibility.  The physical meaning of
group velocities greater than the vacuum speed of light $c$ has been
investigated in experimental work on the propagation of photons
through layered dielectric materials with photonic bandgaps at the
frequency of the light \cite{STE93,STE95,SPI94}.  The measured delays
of individual photons in these experiments were consistent with the
group delay calculated as the derivative of the phase of the
transmission amplitude with respect to angular frequency.  The group
delay can be negative, seemingly consistent with propagation at a
speed greater than $c$.  As has been pointed out previously these
anomalous delays are the result of pulse reshaping and they do not
imply a violation of Einstein causality \cite{STE93,STE95,SPI94}.  

In this paper we take advantage of the regularity of layered
dielectric mirrors to explore explicit effects of pulse reshaping in
the time domain on idealized model pulses, and we use this framework
to interpret the origin of the observed pulse delays.  We study
reshaping of both classical field pulses and fully-quantized
single-photon ``pulses.''  We make a quantitative connection between
the time-domain reshaping and the conventional group delay that yields
insight into the origin of observed negative delays. (Experimentally,
the time-domain detection of ``superluminal'' optical pulses is
difficult, but such studies are possible in the terahertz \cite{REI01}
and microwave \cite{MOJ00} regions of the electromagnetic spectrum, and
analogous acoustic waves that travel faster than the speed of sound are 
experimentally accessible \cite{ROB02}.)  Our time-domain analysis is
a specific example of the ``interference between \dots causally
propagating consecutive components'' that was proposed as universal
mechanism for understanding delays in nondissipative media in
Ref.~\cite{JAP96}, and we extend the classical ideas of
Ref.~\cite{JAP96} into the domain of quantum mechanics and the
transmission of single photons.  Our analysis is complementary to analyses in
the frequency domain like that of Ref.~\cite{ROM01}.

We consider propagation through mirrors comprised of layers of
non-dispersive linear dielectric materials, each of which can be
characterized by a single real index of refraction $n_i$.  This means
that within a single material electromagnetic waves propagate with
equal phase and group velocities given by $c/n_i$.  The reduced speeds
within such materials are partially responsible for observed pulse
delays, but the reshaping due to multiple reflections also plays a
significant role. Analogous anomalous sound speeds due to pulse
reshaping in one-dimensional acoustic bandgaps have also been
investigated experimentally \cite{ROB02}.

We study propagation of plane waves which are normally incident on
layered dielectrics; the thickness of a layer is given by $d_i$ and
the index by $n_i$. To keep things simple we limit the analysis to
materials in which the optical path length is the same for all layers,
which means that the time it takes light to cross a layer is the same
for all layers.  For convenience we define a time $\delta_A$ which is
the round trip time within a layer, i.e., twice the transit time,
\begin{equation}
\delta_A = \frac{2d_in_i}{c},\
\label{eq_delta_A}
\end{equation}
which is independent of the index $i$.
We also define a time $\delta_B$ which corresponds to the delay of a 
wavefront propagating through the entire mirror:
\begin{equation}
\delta_B = \frac{1}{c}\sum_i d_i(n_i-1).
\label{eq_delta_B}
\end{equation}

In Sec.\ref{sec_classical} we examine the propagation of classical
fields and delineate the dependence of the observed delays on the two
parameters $\delta_A$ and $\delta_B$. In Sec.\ \ref{sec_quantum} we
extend the analysis to fully-quantized single photon fields, and draw 
parallels between the propagation of classical and quantum fields.

\section{Propagation of Classical Fields}
\label{sec_classical}

Pulse reshaping in layered dielectrics is most dramatic for pulses
with abrupt changes in amplitude, and we first consider the effects on
square pulses.  Although the reshaping of such pulses leads to
significant distortion, the simplicity of square pulses makes it easy
to disentangle the effects leading to transit-time delays.  After
demonstrating the nature of the reshaping effects with square pulses
we discuss slowly varying pulses, and derive quantitative expressions
for delays.

\subsection{Reshaping of Square Pulses}

We consider a classical square pulse with complex magnitude ${\cal E}$
and duration $\tau$, and assume that in the absence of any dielectric
material this pulse arrives at the observation point at $t=0$, so that
the observed complex field amplitude after propagating through a
vacuum can be written
\begin{equation}
E_1(t) = \left\{\begin{array}{ll}
			0 & \mbox{for $t< 0$} \\
			{\cal E}e^{-i\omega t}&\mbox{for $0\leq t\leq \tau$}\\
			0 & \mbox{for $t>\tau$}
	     \end{array} \right. .
\label{eq_ei}
\end{equation}

When the field passes through a layered dielectric slab the field
arriving at the observation point is composed of multiple reflections.
The leading edge of the transmitted field undergoes no reflections,
and is delayed by the time $\delta_B$; reflected fractions of the
field will be additionally delayed by multiples of $\delta_A$.  We
write the total field arriving at the observation point as the sum of
terms grouped according the how much time the field has spent
traversing the mirror.  The first term corresponds to light that
suffers no reflections; the second to light that spends an ``extra''
time $\delta_A$ within the mirror; the third to light that spends an
``extra'' time $2\delta_A$ within the mirror, etc. (The terms are {\em
not} grouped by the number of reflections.  For example, some fields
spending an ``extra'' time $2\delta$ will have undergone two internal
reflections, while some will have undergone four.)  We separate the
transmission coefficient for each {\em term} into a real factor
describing the attenuation, and a complex factor giving the phase
shift, so that the total field arriving at the observation point is
written
\begin{eqnarray}
E_2(t) &=& {\cal E}e^{-i\omega t}
   \left[\Theta(t-\delta_B)T_0e^{i2\pi c\delta_B/\lambda} \right. \nonumber \\
  && \left. +\Theta(t-\delta_B-\delta_A)T_1 
     e^{i2\pi c(\delta_B+\delta_A)/\lambda}
			\right. \nonumber \\
  && \left. +\Theta(t-\delta_B-2\delta_A)T_2
		e^{i2\pi c(\delta_B+2\delta_A)/\lambda} \right. \nonumber \\
  && + \left.\cdots\right],
\label{eq_et_gen}
\end{eqnarray}
where in this expression (and throughout this article) $\lambda$ refers
to the wavelength in vacuum.  For a single dielectric layer the
attenuation coefficients are
\begin{equation}
T_j =\frac{4n}{(n+1)^2}\left(\frac{n-1}{n+1}\right)^{2j-2}.
\end{equation}
For more complex materials the coefficients $T_i$ are constructed from
products of the appropriate single-interface reflection and
transmission coefficients.

\begin{figure}[t]
\includegraphics{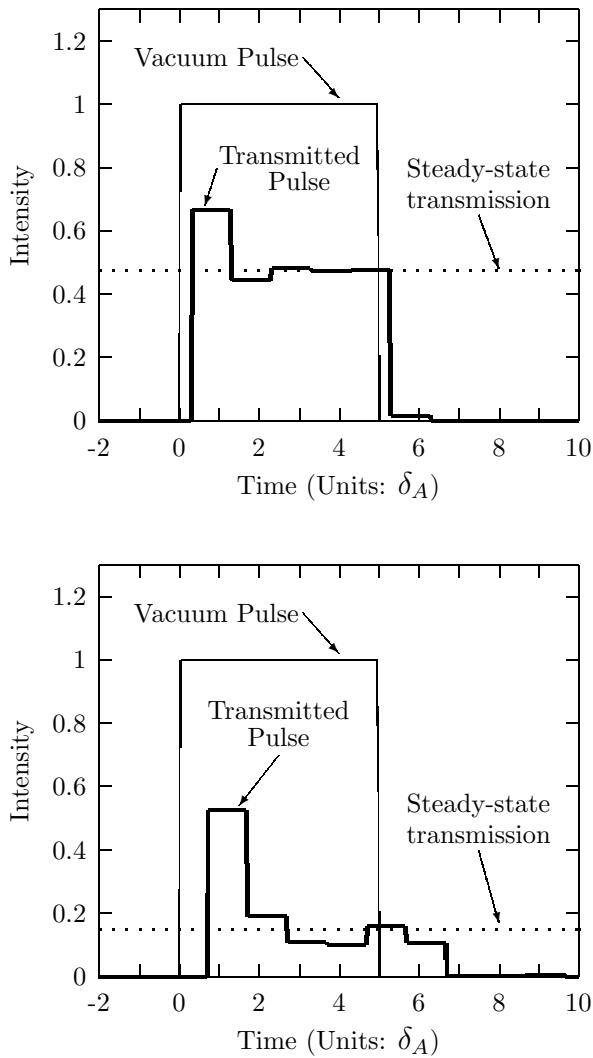}
\caption{Transmission of square pulse through simple dielectric slabs.
The top graph is for a single layer with $n=2.5$ and the bottom is for
three layers with $n_1=2.5$, $n_2=1.25$, and $n_3=2.5$. In both graphs
the optical path length of the dielectric layers is chosen to be
$(m+1/4)\lambda$, where $m$ is an integer, so that the steady-state
transmission is a minimum.  The time units correspond to the
round-trip time through a single layer, i.e., $\delta_A=2n_id_i/c$.
The corresponding leading edge delays, $\delta_B$, are 0.3 and 0.7 in
the upper and lower graphs respectively.}
\label{f_1-3slab_intensity}
\end{figure}

For pulses with durations $\tau$ that are large compared to the
round-trip time $\delta_A$, many terms in Eq.~(\ref{eq_et_gen}) have
time to ``turn on'' before the pulse completely passes by the
observation point.  In the middle of such pulses the effective total
transmission coefficient is
\begin{equation}
T_{\rm total} = \sum_j^N T_j e^{i2\pi c(\delta_B + j\delta_A)},
\label{eq_1slab_steady}
\end{equation}
where $N$ is the number of terms that are ``on'' at the observation
time. As $N$ grows large the transmission approaches that given by the
standard steady-state transmission coefficient, which
can be expressed as
\begin{equation}
T_{\rm s.s.} = \sum_j^\infty T_j e^{i2\pi c(\delta_B + j\delta_A)}.
\label{eq_t_ss}
\end{equation}

Some examples of the effects of simple dielectric layers on square
pulses are illustrated in Fig.~\ref{f_1-3slab_intensity}.  The top
graph in the figure illustrates the time-dependence of the
transmission through a single layer whose optical path length is
$(m+1/4)\lambda$, so that $e^{i2\pi c\delta_A/\lambda}=-1$,
corresponding to a minimum in the steady-state transmission.  The
first arrival of the transmitted pulse is delayed by $\delta_B$, and
this early arriving field is larger in magnitude than the rest of the
pulse because at this time there are not yet any interference effects
reducing the field.  The lower graph displays transmission through a
three-layer mirror.  The additional layers result in an increased
delay in the arrival of the leading edge and a reduction in the
steady-state transmission coefficient.  The additional layers also
result in an enhancement of the large magnitude of the leading-edge
(when compared to the steady-state transmission).  The relatively
large early-time transmission lasts for times on the order of
$\delta_A$, the time between arrival of fields corresponding to terms
in Eq.~(\ref{eq_et_gen}).  Note, though, that for multi-layer mirrors
it takes more time for the effects of all of the multiple reflections
to ``turn on'' at the observation point, and it takes longer for the
transmitted pulse to settle down to the steady-state intensity.

The net effect of the layered dielectric is to create a transmitted
field whose first arrival is delayed by $\delta_B$, but whose temporal
intensity profile is reshaped so that there is relatively more
intensity at early times compared to the intensity profile of the
vacuum pulse.  This shifting of intensity to relatively earlier times
contributes to an effective advance of the pulse.  If a single arrival
time is to be assigned to a pulse (or a photon), the relative role of
these two effects must be accounted for when understanding seemingly
anomalous velocities.

We note that for wavelengths corresponding to transmission maxima, the
intensity profiles would differ from those illustrated in
Fig.~\ref{f_1-3slab_intensity}.  The time of first arrival would be
unchanged, but the initial field would be relatively low, and would
take time to ``build up'' to its steady-state value.  The resulting
pulse profile would show relatively more intensity at later times
compared to the profile of the vacuum pulse, contributing to an
effective delay.

\subsection{Transmission of slowly varying pulses}

For smoothly varying pulses any reshaping effects will be much smaller
than those illustrated for square pulses in the previous section, but
the transmitted pulse is built up from multiple reflections in much
the same way. For pulses that vary slowly enough, the transmitted
pulse that is constructed in this way will have the same shape as the
incident pulse. We demonstrate that the transmitted pulse may be
constructed such that it is delayed or advanced relative to the vacuum
pulse.  (No violation of causality is implied, and the intensity of
the transmitted pulse is always lower than that of the vacuum pulse.)

For simplicity we consider a portion of an incident pulse with linear
amplitude modulation, and assume that the linear modulation has been
in effect since a time $t_0$, so that the incident field can be
written
\begin{equation}
E_1(t) = {\cal E}_0\left[1 + m(t-t_0)\right]e^{-i\omega t}.
\label{eq_field_inc}
\end{equation}
In the following analysis we assume that the round trip time within
the slab is much less than the time that the modulation has been in
effect, i.e., $\delta_A\ll(t-t_0)$.  This means that a very large
number of the terms in a series like that of Eq.~(\ref{eq_et_gen})
have ``turned on,'' and the transmitted field is

\begin{widetext}
\begin{eqnarray}
E_2(t) &=& {\cal E}_0e^{-i\omega t}
   \left\{\left[1 + m(t-t_0-\delta_B)\right]T_0e^{i2\pi c\delta_B/\lambda}
       +\left[1 + m(t-t_0-\delta_B-\delta_A)\right]T_1 
		e^{i2\pi c(\delta_B+\delta_A)/\lambda}
			\right. \nonumber \\
  && \left. + \left[1+m(t-t_0-\delta_B-2\delta_A)\right] T_2
		e^{i2\pi c(\delta_B+2\delta_A)/\lambda}  + \cdots\right\} 
			\nonumber \\
  &=& {\cal E}_0e^{-i\omega t}\sum_{j=0}^N\left[ 1 + 
		m(t-t_0-\delta_B - j\delta_A)\right]T_j
		e^{i2\pi c(\delta_B+j\delta_A)/\lambda},
\end{eqnarray}
where $N$ is the number of terms that have ``turned on.''  When $N$ is 
large, the transmitted field is approximately
\begin{equation}
E_2(t) \simeq   {\cal E}_0e^{-i\omega t}T_{\rm s.s.}
		\left[
		1 + m\left(t-t_0-\delta_B - \delta_A
           \frac{\sum_{j=1}^\infty jT_je^{i2\pi c(\delta_B+j\delta_A)/\lambda}}
			{T_{\rm s.s.}}
			\right)\right].
\end{equation}
To first order in the small quantity $\delta_A/(t-t_0-\delta_B)$ the 
transmitted field is
\begin{equation}
E_2(t) \simeq {\cal E}_0e^{-i(\omega t+\alpha)}T_{\rm s.s.}
	\left\{ 1 + m\left[t-t_0-\delta_B - \delta_A \mbox{Re}
     \left(\frac{\sum_{j=1}^\infty jT_je^{i2\pi c(\delta_B+j\delta_A)/\lambda}}
		{T_{\rm s.s.}}\right)\right]\right\},
\label{eq_et_gen_lin}
\end{equation}
\end{widetext}
where $\alpha$ is a phase shift that will not be of further consequence
in this analysis.  

Comparing the expression for the transmitted field given by 
Eq.~(\ref{eq_et_gen_lin}) to that of the incident field,
Eq.~(\ref{eq_field_inc}), shows that the net result of the multiple
reflections is an effective time delay of the linearly changing field
given by
\begin{equation}
\delta_{\rm effective} = \delta_B + \delta_A \mbox{Re}
    \left(\frac{\sum_{j=1}^\infty jT_je^{i2\pi c(\delta_B+j\delta_A)/\lambda}}
		{T_{\rm s.s.}}\right).
\label{eq_delay_effective}
\end{equation}

This effective delay can be positive, corresponding to a true delay,
or it can be negative, corresponding to an advance.  The exact value
of the delay depends on the indices of refraction in the material
comprising the mirror and the number of layers in the mirror.

The first term in Eq.~(\ref{eq_delay_effective}), $\delta_B$, is
simply the delay due to the reduced speed of wave-fronts within the
dielectric materials. It is always a positive quantity, corresponding
to an actual delay.  The second term contains the more complicated
effects of phased reflections, and may be positive or negative.  If it
is negative and greater in magnitude than $\delta_B$, the multiple
reflection effects that led to the reshaping of square pulses dominate
over the effect of reduced wave-front velocity, and the total delay is
negative.  It is important to note that the effective time delay does
not arise from the simple shifting of the incident field at a given
time to a new time. Rather, the effective delay is the result of the
superposition of attenuated and phase-shifted fields from many
previous times.

Pulses will maintain their shapes and exhibit delays given by
$\delta_{\rm effective}$ as long as the time scale characterizing the
modulation is long compared to $\delta_A$, the time between the
arrival of successive reflections.  The peak of the transmitted pulse
may arrive after the peak of the vacuum pulse would have arrived, or
before.  This is because the transmitted pulse is not the result of
simple attenuation of the incident peak, but rather it is constructed
from the superposition of many reflections, as in
Eq.~(\ref{eq_et_gen}).  Slowly varying pulses are special in the sense
that the newly constructed pulse has the same shape as the incident
pulse.  We emphasize that the effective delay does not apply to the
arrival time of any feature associated with an abrupt change in the
field; the arrival of the leading edge of any disturbance associated
with such an abrupt change will be delayed by $\delta_B$.

We conclude this section by demonstrating that the effective delay
given by Eq.~(\ref{eq_delay_effective}) is identical to that predicted
by the conventional group delay, which is the derivative of the phase
of the transmission amplitude with respect to angular
frequency. Experimental measurements of delays have been consistent
with the group delay, and our time-domain approach gives a physical
picture of the origin of the observed delays.

The steady-state transmission coefficient is given by Eq.~({\ref{eq_t_ss}),
and the phase of this transmission coefficient is
\begin{eqnarray}
\phi &=& \arctan\left[\frac{\mbox{Im}(T_{\rm s.s.})}{\mbox{Re}(T_{\rm s.s.})}
           \right] \nonumber \\
   &=& \arctan\left[\frac{\mbox{Im}
    (\sum_j^\infty T_je^{i2\pi c(\delta_B + j\delta_A)})}
      {\mbox{Re}(\sum_j^\infty T_je^{i2\pi c(\delta_B + j\delta_A)})}
           \right].
\end{eqnarray}
It is straightforward to show that 
\begin{eqnarray}
\frac{d\phi}{d\omega}  &=& \frac{1}{\left[\left(\frac{\mbox{Im}(T_{\rm s.s.})}
  {\mbox{Re}(T_{\rm s.s.})}\right)^2 +1\right]}\frac{d}{d\omega}
  \left[\frac{\mbox{Im}(T_{\rm s.s.})}{\mbox{Re}(T_{\rm s.s.})}\right]
                      \nonumber \\
                &=& \delta_B - \delta_A \mbox{Re}
     \left(\frac{\sum_{j=1}^\infty jT_je^{i2\pi c(\delta_B+j\delta_A)/\lambda}}
		{T_{\rm s.s.}}\right),
\end{eqnarray}
which is equal to $\delta_{\rm effective}$, the effective delay given
by Eq.~(\ref{eq_delay_effective}), which was derived above from
time-domain considerations.  

The equivalence (for slowly varying pulses) of the group delay with
the effective delay derived in the time-domain is a further
demonstration that the group delay has a physical meaning, even in
cases in which it results in a seemingly anomalous advance in the peak
of the transmitted pulse relative to the peak of a vacuum pulse.  For
pulses which vary rapidly on the time-scale given by $\delta_A$ there
will be significant distortion of the shape of the pulse that will
depend on the details of the pulse shape and the characteristics of
the dielectric mirror.  Any discussion of delays for such pulses must
carefully account for such distortions in a way that is beyond the
scope of the present analysis.

\section{Propagation of Single-Photon Quantum Fields}
\label{sec_quantum}

In the preceding section we discussed the effects of layered
dielectric materials on the propagation of classical fields, but the
experimental determinations of propagation time cited above involved
the detection of individual photons.  In this section we demonstrate a
way in which the time-domain picture presented for classical fields
can be extended to single-photon quantum fields.  We develop a model
in which an excited atom spontaneously emits a quantized multimode
photon, and we investigate the time-dependent probability for
excitation of a detector atom located on the opposite side of a
dielectric mirror from the emitting atom.  The time dependent
excitation probability displays interference effects that are exact
analogs to those experienced by classical fields.

\subsection{Quantum Model}

We consider a large one-dimensional multimode optical cavity of total
length $L$ which contains a symmetric dielectric mirror in the center
of the cavity.  (See Fig.~\ref{f_model}.) The mirror is comprised of
layers of homogeneous linear dielectric like that considered in
Sec.~\ref{sec_classical}.  We find the classical standing-wave modes
of the electromagnetic field in a cavity that {\em includes} the
dielectric material, and quantize the modes of this inhomogeneous
cavity.
\begin{figure}[t]
\includegraphics{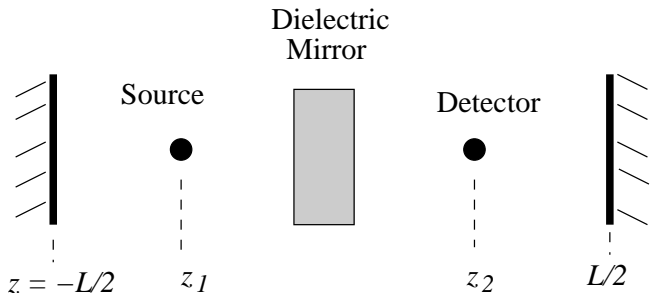}
\caption{Quantum model consisting of a pair of two-level atoms at
fixed positions in a large multimode optical cavity.  A dielectric
mirror is centered in the cavity. We quantize the modes of the {\em
inhomogeneous} cavity. The atom on the left begins in the excited
state and is the source of a photon; the atom on the right serves as a
detector.}
\label{f_model}
\end{figure}

The cavity also contains a pair two-level atoms at fixed positions
$z_1$ and $z_2$ on opposite sides of the dielectric region.  The atom
at $z_1$ is initially in the excited state and spontaneously emits a
photon into the quantized modes of the cavity; it serves as the source
of the quantized field to be transmitted through the mirror.  The atom
at $z_2$ begins in the ground state and serves as a detector of the
transmitted radiation.  (It is also possible to determine equivalent
information about the transmitted radiation from quantities involving
field operators such as the expectation value of the intensity
operator, $\langle \hat{E}^-\hat{E}^+\rangle$ \cite{MAN95}.  We use
the excitation of a two-level atom because of the ease with which can
compute this quantity to high precision in our model.)

The cavity is assumed to be large, in the sense that the length $L$ is
very much greater than the wavelength of the light emitted by the
atoms, i.e., this is {\em not} a microcavity. The finite length $L$
does not contribute to the physical phenomena under investigation; it
simply provides a convenient quantization volume for the field modes
used in our calculations.

We use a standard Hamiltonian of quantum optics to calculate the time
evolution of the system, and pay particular attention to the amplitude
for the atom on the right side of the cavity to be found in the
excited state. We note that the effects of the spontaneously emitted
photon propagate causally in this model.  The explicit form of the
Hamiltonian we use is \cite{MEY99,SAR74,LOU83}
\begin{eqnarray}
H &=& H_{\rm atoms} + H_{\rm field} + H_{\rm interaction} \nonumber \\
  &=& \sum_j\hbar\omega^{\rm (at)}\sigma_z^{(j)} 
       + \sum_m \hbar\omega_m a_m^\dagger a_m  \nonumber \\
  &&   + \sum_{j,m} \hbar \left(g_{jm} a_m \sigma_+^{(j)} + 
         g_{jm}^\ast a_m^\dagger \sigma_-^{(j)}\right), \label{ham}
\end{eqnarray}
in which the atoms are labeled with the index $j$ and the field modes
with index $m$, and where $\omega^{(at)}$ is the zero-field resonance
frequency of both atoms, $\omega_m$ is the frequency of the $m^{\rm
th}$ field mode, $a_m$ and $a_m^\dagger$ are the annihilation and
creation operators for the $m^{\rm th}$ mode, $\sigma_+^{(j)}$,
$\sigma_-^{(j)}$, and $\sigma_z^{(j)}$ are the pseudo-spin operators
which act on atom $j$, and $g_{jm}$ gives the coupling of the $j^{\rm
th}$ atom to the $m^{\rm th}$ field mode.  In this Hamiltonian we have
made the standard electric-dipole and rotating-wave approximations.

The mode frequencies $\omega_m$ in the Hamiltonian are simply those of
the classical standing wave modes of the electromagnetic field.  The
spatial mode functions are normalized so that the energy per photon in
the quantized modes is $\hbar\omega_m$.  The relative magnitudes of
the coupling constants $g_{jm}$ reflect the spatial dependence of the
classical mode functions, specifically the relative magnitude of the
mode functions at the positions of the two atoms.  Calculation of the
mode frequencies and spatial mode functions for the inhomogeneous
cavity involves the solution of a classical boundary value problem.
This is a straightforward process in principle, although the large
number of boundaries in a multi-layer dielectric mirror leads to
algebraic complexity.  We used transfer matrix methods \cite{WAL92}
(adapted to electromagnetic standing waves) and the ``shooting
method'' \cite{PRE88} to determine numerically the mode frequencies.

In the limit of a large cavity we may assume that the frequencies of
all atomic transitions are very much greater than the fundamental
frequency of the cavity. In this limit we can make the approximation
that all modes that influence the dynamics of the system are near the
atomic resonance, and the atom-field coupling constants can be
factored into a product of a frequency-independent constant and a
space-dependent coupling factor.  The coupling constants $g_{jm}$ are
given in terms of the electric dipole matrix element $d_j$ between the
two levels of atom $j$, the effective volume of the cavity $V$, a
mode-dependent normalization factor ${\cal N}_m$, and the permittivity
of free space $\epsilon_0$, by
\begin{eqnarray}
g_{jm} &=& \pm d_j{\cal N}_m\left(\frac{\omega^{\rm (at)}}
                   {2\hbar\epsilon_0 V}\right)
                \sin\left[k_m(L/2-z_j)\right] \nonumber \\
       &=& \pm\Omega_j{\cal N}_m\sin\left[k_m(L/2-z_j)\right],
\label{eq_coupling}
\end{eqnarray}
where $k_m$ is the wave-vector for mode $m$, and in the last line we
have defined the quantity
\begin{equation}
\Omega_j =d_j\left(\frac{\omega^{\rm (at)}_j}{2\hbar\epsilon_0 V}\right)^{1/2},
\end{equation}
which is independent of the cavity mode-frequency.  For symmetrically
place atoms, modes with even spatial mode functions yield coupling
constants with the same sign for each atom; odd mode functions give 
coupling constants of opposite signs.  In performing our numerical 
calculations we use an equal number of modes above and below the
atomic resonance frequency.

The basis states for describing the system are 
\begin{itemize}
\item $\vert e,g;0\rangle$: left atom excited, right atom in ground 
state, no photon,
\item $ \vert g,e;0\rangle$: right atom excited, left atom in ground 
state, no photon,
\item $\vert g,g;1_m\rangle$: both atoms in ground state, one photon in 
$m^{\rm th}$ cavity mode,
\end{itemize}
and we write the state of the system as the linear combination
\begin{eqnarray}
\vert \psi(t)\rangle &=& c_1(t)\vert e,g;0\rangle + c_2(t)\vert g,e;0\rangle
                \nonumber \\
        &&      +\sum_m b_m(t)\vert g,g;1_m\rangle.
\label{eq_psi_gen}
\end{eqnarray}
In all the examples in this paper the system starts in the 
state
\begin{equation}
\vert \psi(0)\rangle = \vert e,g;0\rangle,
\end{equation}
and we pay particular attention to the complex amplitude $c_2(t)$ for
the detector atom to be found in the excited state.  Although it is
difficult physically to prepare a state which corresponds to our
initial condition, this idealized state has the advantage that at
$t=0$ all of the energy is localized at a single point (the position
of the excited atom), making causal wavefronts evolving from this
state particularly easy to identify. (A visualization of the
propagation of the wavefronts of the intensity of the quantum field in
similar models is presented in \cite{LIG02,BUZ99}.)

\subsection{Method of Solution}
\label{sec_method}
It is possible to find analytical solutions for the time evolution of
atom-cavity systems with a single photon in simple inhomogeneous
cavities \cite{LIG02b}, but the complexity of the mode structure for a
cavity with a many-layered mirror makes this approach intractable.
Therefore we construct numerical solutions for the coefficients
$c_1(t)$, $c_2(t)$ and $b_k(t)$ of Eq.~(\ref{eq_psi_gen}).

We use the time-independent Schr\"{o}dinger equation to determine the
energies $E_q$ and eigenstates $\vert E_q\rangle$ of the {\em total}
Hamiltonian.  The time evolution of the system is then straightforward
to calculate.  If the system begins in state
\begin{eqnarray}
\vert \psi(0)\rangle &=& \vert e,g;0\rangle \nonumber \\
                     &=& \sum_q \vert E_q\rangle\langle E_q\vert e,g;0\rangle,
\end{eqnarray}
then the state of the system at a later time $t$ is given by
\begin{equation}
\vert \psi(t)\rangle = \sum_q e^{-iE_qt/\hbar}
                        \vert E_q\rangle\langle E_q\vert e,g;0\rangle.
\label{eq_psi_t_gen}
\end{equation}    
Projecting Eq.~(\ref{eq_psi_t_gen}) onto the basis states gives the 
the time-dependent coefficients of  Eq.~(\ref{eq_psi_gen}):
\begin{eqnarray}
c_1(t) &=&  \langle e,g;0\vert \psi(t)\rangle \nonumber \\
        &=& \sum_q e^{-iE_qt/\hbar}
                \langle e,g;0 \vert E_q\rangle\langle E_q\vert e,g;0\rangle
                        \nonumber \\
        &=& \sum_q e^{-iE_qt/\hbar}
                \left\vert\langle e,g;0 \vert E_q\rangle\right\vert^2,
\end{eqnarray}
\begin{eqnarray}
c_2(t) &=&  \langle g,e;0\vert \psi(t)\rangle \nonumber \\
        &=& \sum_q e^{-iE_qt/\hbar}
                \langle g,e;0 \vert E_q\rangle\langle E_q\vert e,g;0\rangle,
\end{eqnarray}
and 
\begin{eqnarray}
b_k(t) &=&  \langle g,g;1_k\vert \psi(t)\rangle \nonumber \\ 
        &=& \sum_q e^{-iE_qt/\hbar}
                \langle g,g;1_k\vert E_q\rangle\langle E_q\vert e,g;0\rangle.
\end{eqnarray}

We use standard numerical matrix diagonalization routines to determine
the eigenvalues and eigenvectors used in these equations.  We consider
systems with as many as 2,000 modes, which leads to large matrix
representations of the Hamiltonian, but the the matrix is sparse.  Our
approach is similar to that used previously in several studies
\cite{LIG95,LIG02,BUZ99}.

\subsection{Reshaping of single photon ``pulses''}
\label{sec_results}

The graphs of this section demonstrate the analogy between the
classical field and the quantum amplitude to find the detector atom in
the excited state.  The total quantum amplitude is the result of
interference from multiple reflections in very much the same way as
the classical field is the result of multiple reflections.  In this
section we present results for three cavities: an empty, or vacuum
cavity, a cavity with a simple homogeneous dielectric region, and a
cavity containing a dielectric mirror.  The results for the simple
cavity help elucidate the more complicated behavior seen in the
propagation of through the dielectric mirror.

For atoms in an empty cavity with no dielectric mirror it is possible
to find an analytic solution for the dynamics \cite{PUR02a}.  The
probability for the source atom to be in the excited state decays
exponentially with decay constant $\gamma_1=\vert\Omega_1\vert^2 L/c$
until the time at which reflections first interrupt the decay; for
atomic positions $z_1=-.25L$ and $z_2=.25L$ this decay proceeds until
reflections return to the atom time $t=0.5L/c$.  For atoms at these
positions the amplitude to find the detector atom in the excited state
remains identically zero until time $t=0.5L/c$, the time at which
radiation can first reach the detector. Choosing units such that
$L/c=1$ (which we will use for the remainder of the article) the
amplitude to find detector atom in the excited state in an empty
cavity is \cite{PUR02a}
\begin{figure}[t]
\includegraphics{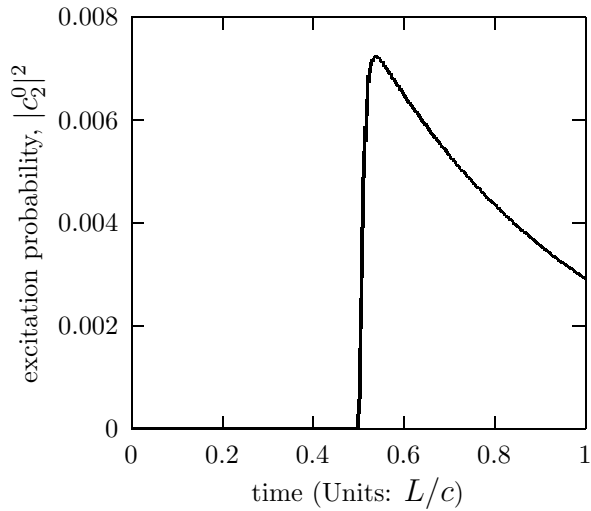}
\caption{Detector atom excitation probability in an empty cavity.  The
source atom decay rate is $\gamma_1=16$ and the detector atom rate is
$\gamma_2=256$.  The rapid rise at $t=1/2$ is due to the arrival of
the field at atom 2.  The rate of the rise is determined by the
detector response, and the slow exponential decay reflects the
exponential shape of the pulse passing the detector atom.}
\label{f_c30}
\end{figure}
\begin{eqnarray}
c^{0}_2(t) &=& \Theta\left(t-\frac{1}{2}\right)
        \frac{\sqrt{\gamma_1\gamma_2}}{\gamma_1-\gamma_2} \nonumber \\
        &&  \times\left(e^{-\frac{\gamma_1}{2}(t-\frac{1}{2})} -
                e^{-\frac{\gamma_2}{2}(t-\frac{1}{2})} \right), 
\label{eq_c30}
\end{eqnarray}
in which the decay constant for the detector atom is $\gamma_2 =
\vert\Omega_2\vert^2 L/c$, and the step-function expresses the turn-on
of the excitation at $t=0.5$ and the causal dynamics inherent in our
model.

In the limit of very fast detector atom response, i.e.,
$\gamma_2\rightarrow \infty$, the amplitude $c_2(t)$ will
instantaneously reflect the strength of the field incident on the
detector atom.  Fig.~\ref{f_c30} illustrates the detector atom
response for the case $\gamma_1=16$ and $\gamma_2=256$.  The rapid
rise reflects the response of the detector atom to the sudden arrival
at $t=0.5$ of radiation from the source atom, and occurs on the time
scale given by $1/\gamma_2$. The slower decline reflects the
exponential shape of the radiation pulse emitted by the source as it
passes by the detector atom \cite{LIG02,BUZ99}.  While in a given run
of an actual experiment the excitation of the detector atom will be
observed at a particular instant of time, the evolution of the
excitation probability is continuous, and resembles the excitation
expected for a classical oscillator driven by a classical field pulse.

An illustration of the effect of a single dielectric slab on a pulse
is given in Fig.~\ref{f_1slab_probability}.
\begin{figure}[b]
\includegraphics{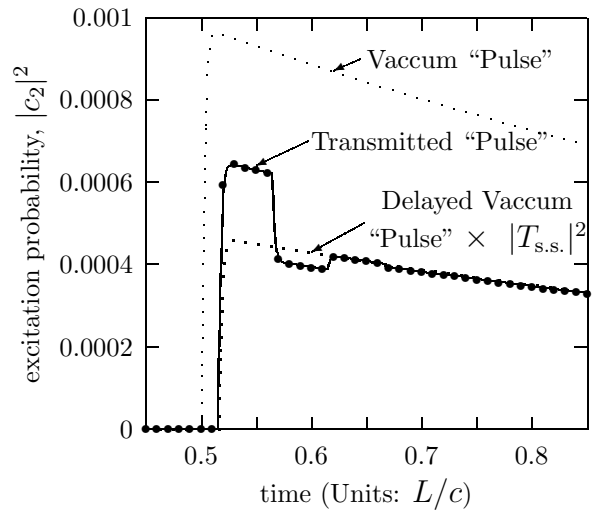}
\caption{Detector atom excitation probability in a cavity with a
single homogeneous dielectric slab.  The source atom decay rate is
$\gamma_1=1$ and the detector atom rate is $\gamma_2=1024$. The width
and index of the slab are chosen so that the classical delays are
$\delta_A = 0.05$ and $\delta_B = 0.015$.  The data points are the
result of fully quantum-mechanical calculations, while the solid line
is built from attenuated, delayed, and phase shifted versions of the
vacuum response using {\em classical} delays, phase-shifts, and
attenuation factors.}
\label{f_1slab_probability}
\end{figure}
This figure gives the time-dependence of the detector atom excitation
probability after transmission of the radiation through a single
dielectric slab whose width and index are chosen so that the classical
delays are $\delta_A = 0.05$ and $\delta_B = 0.015$ in the units of
the figure. Because the decay rate of the source atom is relatively
slow, this figure is analogous to the top graph in
Fig.~\ref{f_1-3slab_intensity}, which gives the intensity of a
classical square pulse after passing through a single slab of
dielectric material.  The excitation of the detector atom ``turns on''
at the expected classical time, $0.5 + \delta_B$, and is interrupted
at multiples of $\delta_A$, the classical round-trip time within the
material.  The excitation probability ``settles down'' to an
attenuated and delayed version of the vacuum ``pulse'' of
Eq.~(\ref{eq_c30}), where the attenuation factor is given by the {\em
classical} $\vert T_{\rm s.s.}\vert^2$.

The data points in the figure are the result of the fully quantum
mechanical calculations described in Sec.~\ref{sec_method}, while the
solid line combines the quantum results for excitation in an empty
cavity with the classical techniques for multiple reflections.  In
Sec.~\ref{sec_classical} the classical field is built up from a sum of
appropriately attenuated, delayed, and phase shifted fields in
Eq.~(\ref{eq_et_gen}); the solid line in
Fig.~\ref{f_1slab_probability} is built up from attenuated, delayed,
and phase shifted versions of the quantum excitation amplitude
Eq.~(\ref{eq_c30}) using the same {\em classical} attenuation, delay,
and phase shift parameters that were used to produce the upper graph
in Fig.~\ref{f_1-3slab_intensity}, i.e.,
\begin{eqnarray}
\mbox{solid line}&\longleftrightarrow& \left[ c_2^0(t-\delta_B)T_0
                e^{i2\pi c\delta_B/\lambda} \right. \nonumber \\
       &&+c_2^0(t-\delta_B-\delta_A)T_1e^{i2\pi c(\delta_A+\delta_B)/\lambda}
		      \nonumber \\
      && +c_2^0(t-\delta_B-2\delta_A)T_2e^{i2\pi c(\delta_A+2\delta_B)/\lambda}
		  \nonumber \\
        &&  + \cdots \Bigr]^2.
\label{eq_c2_semiclassical}
\end{eqnarray}  

\begin{figure}[b]
\includegraphics{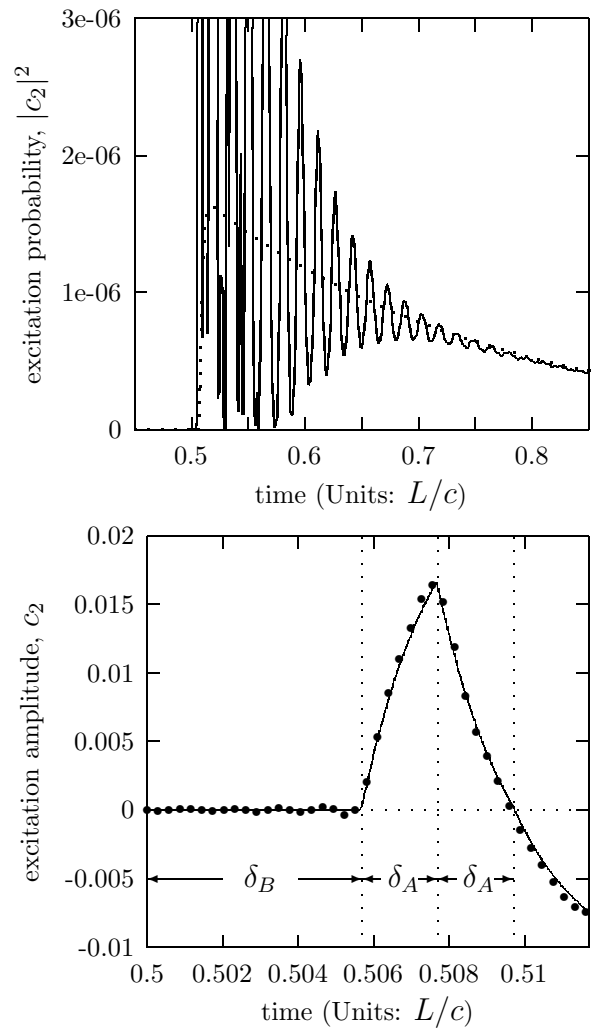}
\caption{Detector atom excitation in a cavity with an (HL)$^5$H
structure dielectric mirror.  The source atom decay rate is
$\gamma_1=4$ and the detector atom rate is $\gamma_2=1024$. The width
of the mirror elements is such that the classical delays are $\delta_A
\simeq 0.002$ and $\delta_B \simeq 0.0057$.  The data points are the
result of fully quantum-mechanical calculations, while the solid line
is built from attenuated, delayed, and phase shifted versions of the
vacuum response using {\em classical} delays, phase-shifts, and
attenuation factors.  See Eq.~(\ref{eq_c2_semiclassical}).}
\label{f_c3_mirror}
\end{figure}
The same principles apply to propagation through more complicated
dielectric mirror structures.  The excitation amplitude after
transmission through an 11 element mirror with alternating high and
low index regions is illustrated in Fig.~\ref{f_c3_mirror}.  This
mirror is similar to the (HL)$^5$H mirror used in the experiments of
references \cite{STE93,STE95}.  The ``high'' index of refraction is
$n_H = 3.0$ and the ``low'' index is $n_L=1.5$, giving an intensity
transmission coefficient at the minimum in the transmission of $\vert
T_{\rm s.s.}\vert^2 = 4.34\times 10^{-4}$.  The upper graph in this
figure also shows the empty-cavity excitation probability given by the
square of Eq.~(\ref{eq_c30}) scaled by the {\em classical}
steady-state transmission factor, $\vert T_{\rm s.s.}\vert^2$.

The effects of the abrupt turn-on of the excitation are evident in the
large variations in the early-time detection probability before the
effects of multiple reflection have taken full effect.  These
variations are analogous to those in Figs.~\ref{f_1-3slab_intensity}
and \ref{f_1slab_probability}, but they last longer in this case
because of the relatively large number of layers comprising the
mirror.  The lower graph in Fig.~\ref{f_c3_mirror} displays the
initial turn-on of the excitation of the detector atom.  (Note that
the vertical scale on this graph corresponds to amplitude rather than
probability; the amplitude can assume negative values just as the
classical field $E_2$ can.)  As in Fig.~\ref{f_1slab_probability}, the
data points are the result of the fully quantum mechanical
calculations, while the solid line combines the quantum results for
excitation in an empty cavity from Eq.~(\ref{eq_c30}) with the
classical techniques for multiple reflections as in
Eq.~(\ref{eq_c2_semiclassical}).  There is no excitation before the
time $t=0.5 + \delta_B$, and at later times the excitation is
interrupted after successive multiples of the single-layer round trip
time $\delta_A$.  We call attention to the vastly different excitation
probabilities at early and late times.  The initial large peaks occur
before the effects of multiple reflection have reduced the
transmission.

The finite rise time evident in the lower graph in
Fig.~\ref{f_c3_mirror} is determined by the response of the detector
atom characterized by $\gamma_2$.  A more rapid detector response
would show a more abrupt rise which more closely follows the
step-function turn-on of the field at the detector atom.  The modeling
of more rapid changes would require the inclusion of more modes in our
numerical analysis.

The transmission of a photon from the source through a dielectric
mirror and to the detector can happen via many indistinguishable
pathways: it can travel directly without undergoing reflection; it may
undergo a single reflection in one of many ways, or it may undergo
multiple reflections.  The numerical results of this section
demonstrate that the excitation amplitude for the detector atom is
built up from interfering amplitudes for all of these processes in
exactly the same way that the the classical transmitted field is built
up from multiple reflections.

\section{Conclusion}

We have examined the transmission of radiation through layered
dielectric mirrors in the time domain.  For slowly varying classical
pulses we have derived a formula for pulse delays that takes into
account the time-domain buildup of the steady-state transmission.  The
finite time that it takes to build up this steady-state transmission
results in pulse reshaping, and our formula helps delineate the
competing effects of pulse reshaping and the reduced front velocities
in dielectric materials.  The delays calculated in our model are
equivalent to those calculated from the conventional group delay, but
our model provides an aid to understanding and interpreting the origin
of the anomalous delay times that have been observed in experiments.

We have also demonstrated that our interpretation can be extended to
include quantized fields.  We have studied numerically a model in
which a spontaneously emitted photon propagates through a layered
dielectric mirror, and excites a detector atom.  The quantum amplitude
to find the atom in the excited state is the result of interfering
terms due to the possibility of multiple reflections in much the same
way that the classical field is the result of the interference of
multiply reflected fields.  The terms in the quantum amplitude
``turn-on'' at exactly the same time as the classical field terms, and
with exactly the same relative amplitudes and phases.  Individual
photons will be detected at a range of times described by this
amplitude, and it is the distribution of arrival times of single
photons that is shifted in time in exactly the same way as the
classical pulse is shifted.

\begin{acknowledgments}
Two of the authors (A\ G.-C. and C.\ D.) acknowledge support from
National Science Foundation Research Experiences for Undergraduates
Program (Grant Number PHYS-9732158).
\end{acknowledgments}

%\bibliography{/home/mligare/research/research.bib}

\begin{thebibliography}{19}
\expandafter\ifx\csname natexlab\endcsname\relax\def\natexlab#1{#1}\fi
\expandafter\ifx\csname bibnamefont\endcsname\relax
  \def\bibnamefont#1{#1}\fi
\expandafter\ifx\csname bibfnamefont\endcsname\relax
  \def\bibfnamefont#1{#1}\fi
\expandafter\ifx\csname citenamefont\endcsname\relax
  \def\citenamefont#1{#1}\fi
\expandafter\ifx\csname url\endcsname\relax
  \def\url#1{\texttt{#1}}\fi
\expandafter\ifx\csname urlprefix\endcsname\relax\def\urlprefix{URL }\fi
\providecommand{\bibinfo}[2]{#2}
\providecommand{\eprint}[2][]{\url{#2}}

\bibitem[{\citenamefont{Steinberg et~al.}(1993)\citenamefont{Steinberg, Kwiat,
  and Chiao}}]{STE93}
\bibinfo{author}{\bibfnamefont{A.~M.} \bibnamefont{Steinberg}},
  \bibinfo{author}{\bibfnamefont{P.~G.} \bibnamefont{Kwiat}}, \bibnamefont{and}
  \bibinfo{author}{\bibfnamefont{R.~Y.} \bibnamefont{Chiao}},
  \bibinfo{journal}{Phys. Rev. Lett.} \textbf{\bibinfo{volume}{71}},
  \bibinfo{pages}{708} (\bibinfo{year}{1993}).

\bibitem[{\citenamefont{Steinberg and Chiao}(1995)}]{STE95}
\bibinfo{author}{\bibfnamefont{A.~M.} \bibnamefont{Steinberg}}
  \bibnamefont{and} \bibinfo{author}{\bibfnamefont{R.~Y.} \bibnamefont{Chiao}},
  \bibinfo{journal}{Phys. Rev. A} \textbf{\bibinfo{volume}{51}},
  \bibinfo{pages}{3525} (\bibinfo{year}{1995}).

\bibitem[{\citenamefont{Spielmann et~al.}(1994)\citenamefont{Spielmann,
  Szip\"{o}cs, Stingl, and Krausz}}]{SPI94}
\bibinfo{author}{\bibfnamefont{C.}~\bibnamefont{Spielmann}},
  \bibinfo{author}{\bibfnamefont{R.}~\bibnamefont{Szip\"{o}cs}},
  \bibinfo{author}{\bibfnamefont{A.}~\bibnamefont{Stingl}}, \bibnamefont{and}
  \bibinfo{author}{\bibfnamefont{F.}~\bibnamefont{Krausz}},
  \bibinfo{journal}{Phys. Rev. Lett.} \textbf{\bibinfo{volume}{73}},
  \bibinfo{pages}{2308} (\bibinfo{year}{1994}).

\bibitem[{\citenamefont{Reiten et~al.}(2001)\citenamefont{Reiten, McClatchey,
  Grischkowsky, and Cheville}}]{REI01}
\bibinfo{author}{\bibfnamefont{M.}~\bibnamefont{Reiten}},
  \bibinfo{author}{\bibfnamefont{K.}~\bibnamefont{McClatchey}},
  \bibinfo{author}{\bibfnamefont{D.}~\bibnamefont{Grischkowsky}},
  \bibnamefont{and} \bibinfo{author}{\bibfnamefont{R.}~\bibnamefont{Cheville}},
  \bibinfo{journal}{Opt. Lett.} \textbf{\bibinfo{volume}{26}},
  \bibinfo{pages}{1900} (\bibinfo{year}{2001}).

\bibitem[{\citenamefont{Mojahedi et~al.}(2000)\citenamefont{Mojahedi,
  Schamiloglu, Hegeler, and Malloy}}]{MOJ00}
\bibinfo{author}{\bibfnamefont{M.}~\bibnamefont{Mojahedi}},
  \bibinfo{author}{\bibfnamefont{E.}~\bibnamefont{Schamiloglu}},
  \bibinfo{author}{\bibfnamefont{F.}~\bibnamefont{Hegeler}}, \bibnamefont{and}
  \bibinfo{author}{\bibfnamefont{K.~J.} \bibnamefont{Malloy}},
  \bibinfo{journal}{Phys. Rev. E} \textbf{\bibinfo{volume}{62}},
  \bibinfo{pages}{5758} (\bibinfo{year}{2000}).

\bibitem[{\citenamefont{Robertson et~al.}(2002)\citenamefont{Robertson, Ash,
  and McGaugh}}]{ROB02}
\bibinfo{author}{\bibfnamefont{W.~M.} \bibnamefont{Robertson}},
  \bibinfo{author}{\bibfnamefont{J.}~\bibnamefont{Ash}}, \bibnamefont{and}
  \bibinfo{author}{\bibfnamefont{J.~M.} \bibnamefont{McGaugh}},
  \bibinfo{journal}{Am. J. Phys.} \textbf{\bibinfo{volume}{70}},
  \bibinfo{pages}{689} (\bibinfo{year}{2002}).

\bibitem[{\citenamefont{Japha and Kurizki}(1996)}]{JAP96}
\bibinfo{author}{\bibfnamefont{Y.}~\bibnamefont{Japha}} \bibnamefont{and}
  \bibinfo{author}{\bibfnamefont{G.}~\bibnamefont{Kurizki}},
  \bibinfo{journal}{Phys. Rev. A} \textbf{\bibinfo{volume}{53}},
  \bibinfo{pages}{586} (\bibinfo{year}{1996}).

\bibitem[{\citenamefont{Romero-Roch\'{i}n
  et~al.}(2001)\citenamefont{Romero-Roch\'{i}n, Duarte-Zamorano,
  Nilsen-Hofseth, and Barrera}}]{ROM01}
\bibinfo{author}{\bibfnamefont{V.}~\bibnamefont{Romero-Roch\'{i}n}},
  \bibinfo{author}{\bibfnamefont{R.~P.} \bibnamefont{Duarte-Zamorano}},
  \bibinfo{author}{\bibfnamefont{S.}~\bibnamefont{Nilsen-Hofseth}},
  \bibnamefont{and} \bibinfo{author}{\bibfnamefont{R.~G.}
  \bibnamefont{Barrera}}, \bibinfo{journal}{Phys. Rev. E}
  \textbf{\bibinfo{volume}{63}}, \bibinfo{pages}{027601}
  (\bibinfo{year}{2001}).

\bibitem[{\citenamefont{Mandel and Wolf}(1995)}]{MAN95}
\bibinfo{author}{\bibfnamefont{L.}~\bibnamefont{Mandel}} \bibnamefont{and}
  \bibinfo{author}{\bibfnamefont{E.}~\bibnamefont{Wolf}},
  \emph{\bibinfo{title}{Optical coherence and quantum optics}}
  (\bibinfo{publisher}{Cambridge University Press},
  \bibinfo{address}{Cambridge}, \bibinfo{year}{1995}).

\bibitem[{\citenamefont{Meystre and Sargent}(1999)}]{MEY99}
\bibinfo{author}{\bibfnamefont{P.}~\bibnamefont{Meystre}} \bibnamefont{and}
  \bibinfo{author}{\bibfnamefont{M.}~\bibnamefont{Sargent}},
  \emph{\bibinfo{title}{Elements of Quantum Optics}}
  (\bibinfo{publisher}{Springer}, \bibinfo{address}{Berlin},
  \bibinfo{year}{1999}), \bibinfo{edition}{3rd} ed.

\bibitem[{\citenamefont{Sargent et~al.}(1974)\citenamefont{Sargent, Scully, and
  Lamb}}]{SAR74}
\bibinfo{author}{\bibfnamefont{M.}~\bibnamefont{Sargent}},
  \bibinfo{author}{\bibfnamefont{M.~O.} \bibnamefont{Scully}},
  \bibnamefont{and} \bibinfo{author}{\bibfnamefont{W.~E.} \bibnamefont{Lamb}},
  \emph{\bibinfo{title}{Laser Physics}} (\bibinfo{publisher}{Addison-Wesley},
  \bibinfo{address}{Reading, MA}, \bibinfo{year}{1974}).

\bibitem[{\citenamefont{Loudon}(1983)}]{LOU83}
\bibinfo{author}{\bibfnamefont{R.}~\bibnamefont{Loudon}},
  \emph{\bibinfo{title}{The Quantum Theory of Light}}
  (\bibinfo{publisher}{Oxford U.~P.}, \bibinfo{address}{Oxford},
  \bibinfo{year}{1983}), \bibinfo{edition}{2nd} ed.

\bibitem[{\citenamefont{Walker}(1992)}]{WAL92}
\bibinfo{author}{\bibfnamefont{J.~S.} \bibnamefont{Walker}},
  \bibinfo{journal}{Comp. Phys.} \textbf{\bibinfo{volume}{6}},
  \bibinfo{pages}{393} (\bibinfo{year}{1992}).

\bibitem[{\citenamefont{Press et~al.}(1988)\citenamefont{Press, Flannery,
  Teukolsky, and Vetterling}}]{PRE88}
\bibinfo{author}{\bibfnamefont{W.~H.} \bibnamefont{Press}},
  \bibinfo{author}{\bibfnamefont{B.~P.} \bibnamefont{Flannery}},
  \bibinfo{author}{\bibfnamefont{S.~A.} \bibnamefont{Teukolsky}},
  \bibnamefont{and} \bibinfo{author}{\bibfnamefont{W.~T.}
  \bibnamefont{Vetterling}}, \emph{\bibinfo{title}{Numerical Recipes in C}}
  (\bibinfo{publisher}{Cambridge University Press},
  \bibinfo{address}{Cambridge}, \bibinfo{year}{1988}).

\bibitem[{\citenamefont{Ligare and Oliveri}(2002)}]{LIG02}
\bibinfo{author}{\bibfnamefont{M.}~\bibnamefont{Ligare}} \bibnamefont{and}
  \bibinfo{author}{\bibfnamefont{R.}~\bibnamefont{Oliveri}},
  \bibinfo{journal}{Am. J. Phys.} \textbf{\bibinfo{volume}{70}},
  \bibinfo{pages}{58} (\bibinfo{year}{2002}).

\bibitem[{\citenamefont{Bu\v{z}ek et~al.}(1999)\citenamefont{Bu\v{z}ek,
  Drobn\'{y}, Kim, Havukainen, and Knight}}]{BUZ99}
\bibinfo{author}{\bibfnamefont{V.}~\bibnamefont{Bu\v{z}ek}},
  \bibinfo{author}{\bibfnamefont{G.}~\bibnamefont{Drobn\'{y}}},
  \bibinfo{author}{\bibfnamefont{M.~G.} \bibnamefont{Kim}},
  \bibinfo{author}{\bibfnamefont{M.}~\bibnamefont{Havukainen}},
  \bibnamefont{and} \bibinfo{author}{\bibfnamefont{P.~L.}
  \bibnamefont{Knight}}, \bibinfo{journal}{Phys. Rev. A}
  \textbf{\bibinfo{volume}{60}}, \bibinfo{pages}{582} (\bibinfo{year}{1999}).

\bibitem[{\citenamefont{Ligare}()}]{LIG02b}
\bibinfo{author}{\bibfnamefont{M.}~\bibnamefont{Ligare}},
  \bibinfo{note}{unpublished paper available at
  http://www.eg.bucknell.edu/phyics/ligare.html}

\bibitem[{\citenamefont{Ligare and Becker}(1995)}]{LIG95}
\bibinfo{author}{\bibfnamefont{M.}~\bibnamefont{Ligare}} \bibnamefont{and}
  \bibinfo{author}{\bibfnamefont{S.~F.} \bibnamefont{Becker}},
  \bibinfo{journal}{Am. J. Phys.} \textbf{\bibinfo{volume}{63}},
  \bibinfo{pages}{788} (\bibinfo{year}{1995}).

\bibitem[{\citenamefont{Purdy et~al.}()\citenamefont{Purdy, Taylor, and
  Ligare}}]{PUR02a}
\bibinfo{author}{\bibfnamefont{T.}~\bibnamefont{Purdy}},
  \bibinfo{author}{\bibfnamefont{D.~F.} \bibnamefont{Taylor}},
  \bibnamefont{and} \bibinfo{author}{\bibfnamefont{M.}~\bibnamefont{Ligare}},
  \bibinfo{note}{arXiv:quant-ph/0204009}

\end{thebibliography}

\end{document}